# Enhanced Fault Ride-Through Grid Forming with Transient Synchronisation Stability and Current Saturation

Youcefa Brahim Elkhalil, Nima Tashakor, Davood Keshavarzi, Ehsan Asadi, and Stefan Goetz

*Abstract*—During grid faults, grid-forming converters are typically suggested to switch from a voltage-source to a current-source mode to limit the current and protect the electronics. This transition has the potential for the converter to transiently lose synchronization due to such current saturation. Therefore, this paper proposes an alternative current saturation algorithm to improve transient synchronization stability during mode switching. The algorithm is designed for grid-forming converters to meet low-voltage ride-through (LVRT) requirements and grid-fault standards in addition to transient synchronization stability. Moreover, it limits the converter output current during grid faults with a new control parameter. The presented method introduces converter output virtual fluxes to calculate the current references in the *d*- and *q*-axes for the current saturation algorithm to enhance LVRT performance and grid stability. The method exploits the correlation between the converter's virtual fluxes and currents to modify the current saturation levels through real-time converter virtual flux estimation. The adaptive saturation levels ensure precise control and high dynamics during grid faults and facilitate optimal power injection or absorption to support the grid. The proposed current-saturation algorithm is analytically evaluated. Further, hardware-in-the-loop (HIL) experiments validate the effectiveness of the proposed algorithm.

## I. INTRODUCTION

Distributed generators, such as solar installations, typically feed electricity into the grid through voltage-source converters [1], [2]. Traditionally, distributed generators are regulated as current sources that synchronize to the grid voltage through a phase-locked loop (PLL), which is called grid-following behavior. Thus, grid-following converters expect a stiff voltage with well-defined frequency provided by some other source, e.g., conventional synchronous generators, and do not control them actively [3]. However, since they nevertheless influence it, a large share of grid-following converters with grid-voltage dominance though without active control of it poses a potential threat to the stable operation of modern power grids. Additionally, PLL can lose synchronization under weak grid conditions [4-7].

To solve challenges related to power system stability and security [8-11], power converters can shift from grid-following to grid-forming control [12]. In grid-forming control, the converter operates as a voltage source and actively regulates the voltage. However, due to the low impedance of grids already under normal conditions and more so in faults, this voltage source behavior necessitates specific attention to over-current protection [13]. Unlike synchronous generators with their massive thermal mass often capable of withstanding up to ten times their rated current, transistors have a very short thermal time constant and are at present chosen very close to the rating for low cost so that power converters can only handle a few percent of over-current (typically some 20%) [14, 15]. Furthermore, with the current shift toward wide-band-gap semi-conductor technologies, the thermal capacity of the power electronic devices can be even further reduced [16]. Consequently, grid-forming converters must be safeguarded against severe faults such as short circuits, voltage sag, heavy load connections, and voltage phase jumps, exclusively relying on control, while maintaining synchronization with the power system.

In the existing literature, current-limiting methods are broadly categorized into virtual impedance (VI) and current saturation algorithms for current limitation [17]. Virtual impedance introduces a pseudo-impedance to restrict the current [18], whereas the current saturation algorithm aims to limit the current reference to the maximum current-carrying capacity of the grid-forming converter [19]. Further, recent research explores the impact of the current phase angle of the converter on the transient stability of grid-forming converters [20]. In addition to a proper current saturation algorithm, the integration of the grid-forming converter into the grid requires adherence to grid codes and low voltage ride-through (LVRT) capability [10, 21]. Furthermore, during grid faults, grid-forming converters are switched from a voltage-source to a current-source mode to enable saturation of the current. This transition has the potential to deteriorate the transient synchronization stability of the grid-forming converter due to the current saturation algorithm.

Previous research extensively studied transient behavior and instability in grid-forming converters with current limiters during mode transitions [22]. The method for assessing transient stability, involving hard limiters represented by equality constraints and numerical solutions, is complex and applies to both grid-following and grid-forming converters [19, 20, 23]. Mode transitions reduce the transient synchronization stability margin of grid-forming converters [23]. Researchers identified the instability mechanism during transient synchronization with current saturation algorithms [19], which allowed them to determine the critical clearance time for grid faults to ensure rapid recovery and a return to stable operation [20].

Methods proposed to enhance transient synchronization stability during mode switching face challenges in practical application due to drawbacks such as high costs for increased system capacity [19]. Furthermore, adding a transient stability enhancement component to the grid-forming control complicates control of power with coupling active and reactive powers [19]. Moreover, using virtual impedance to limit currents for grid forming might result in delayed suppression of over-current [24], despite potential improvements in transient synchronization stability [25], [26], [18]. Therefore, exploring alternative approaches is necessary for enhancing transient synchronization stability during mode change.

Adjusting the current saturation algorithm could offer control over transient synchronization stability. The traditional current saturation algorithm for voltage-source converters, which sets the reference current at a fixed value without quantitative calculations, primarily focuses on equipment protection rather than system stability [27]. However, this strategy has the risk of transient instability during fault conditions caused by mode switching. Previous studies show various transient behaviors in grid-forming converters with different current-saturation methods and suggest that adjusting current limits could improve transient synchronization stability [28], [20]. Dynamically modifying current saturation algorithms could effectively enhance the transient behavior of grid-forming converters. Further research is needed to investigate the enhancement of transient synchronization stability through adjustments of the current saturation algorithm.

The notion of virtual flux, derived from the output current and the integral of the voltage measured at the converter's terminals, has been extensively discussed in the literature [29], [30]. The virtual flux is directly proportional to the internal voltage and provides a straightforward as well as robust means of controlling the converter. Historically, virtual-flux is suggested for converter droop control in microgrids [31], used in direct control for three-phase rectifiers [32], and applied in predictive control scenarios [29]. In contrast to these approaches, the current saturation control method presented in this paper is centered on virtual-flux control, introducing a novel current saturation algorithm for grid-forming converter control.

The method presented in this paper resolves the limitations of conventional current saturation algorithms by introducing a novel calculation of the current phase angle of the converter based on a unique estimation method for the $d$- and $q$-axis current references. Therefore, this angle is determined by estimating virtual fluxes at both the converter and the grid terminals, which are then used to compute the current flowing through the filter inductor, a critical component in grid-forming converter systems. Further, the virtual fluxes are computed through the integration of voltage signals from both converter and grid sides, followed by transformation into $d$ and $q$ components using Park transformation. Furthermore, the integration process inherently acts as a filter in generating new filtered current references with a stable and clean current phase angle. Consequently, we enhance system stability and transient performance, particularly during grid faults. Our approach promises a significant advancement in the field of grid-forming converter control strategies, offering improved system robustness and reliability in challenging grid conditions.

The paper is organized as follows: Sections II and III summarize grid-forming converter modeling and droop control. Section IV examines current saturation for maintaining system stability under conditions of high demand. Section V analyses the stability of droop control of a grid-forming converter to derive a relationship between droop control and system resilience. Section VI introduces a current saturation algorithm to effectively enhance stability. Section VII details the proposed current-phase-angle estimation for the current saturation algorithm and demonstrates how it improves accuracy and performance, while Section VIII summarizes and discusses the results followed by the conclusion in Section IX.

## II. GRID-FORMING CONVERTER MODELING

### A. Fundamental Configuration

Fig. 1 illustrates the system model of a grid-tied grid-forming converter. The converter receives power from a DC voltage source of $V_{dc}$. The LC filter ($R_f$, $L_f$ and $C_f$) connects the converter on the AC side to the grid at the point of common coupling (PCC). The grid itself has an AC voltage source $V_{grid}$ with impedance ($R_g$, $L_g$). Within the same figure, the control section includes a voltage control loop, a current-saturation algorithm, and an inner current-control loop. An external power-control loop facilitates grid synchronization without relying on a phase-locked loop. It operates on $P$–$\omega$ droop for active and $Q$–$v$ for reactive power synchronization.

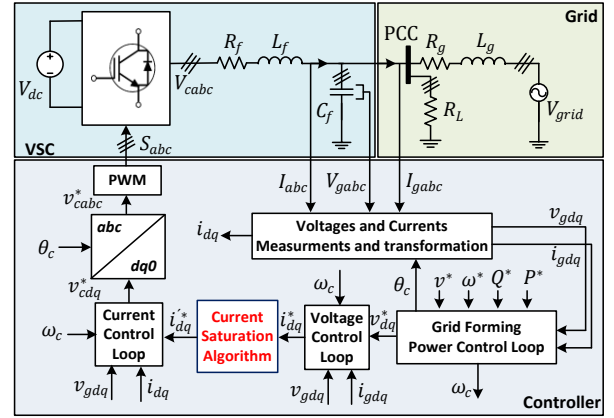

Fig. 1. System control diagram of a grid-tied grid-forming converter with current saturation.

### B. System Modelling

The schematic diagram in Fig. 1 illustrates a three-phase grid-forming system connected to a grid model, which mathematically follows

$$v_{cp} = L_f \frac{di_p}{dt} + R_f i_p + v_{gp}, \quad (1)$$

with $p \in \{a, b, c\}$ denoting the specific phase.

The voltages of the three filter capacitors of the converter, $v_{ga}$, $v_{gb}$, $v_{gc}$ follow the corresponding filter currents $i_a$, $i_b$, $i_c$, and load currents $i_{ga}$, $i_{gb}$, $i_{gc}$ per

$$v_{gp} = \frac{1}{C_f} \int (i_p - i_{gp}). \quad (2)$$

In the $dq0$ frame, Equations (1) and (2) transform into

$$v_{cd} = L_f \frac{di_d}{dt} + R_f i_d + v_{gd} - \omega_c L_f i_q,$$
$$v_{cq} = L_f \frac{di_q}{dt} + R_f i_q + v_{gq} + \omega_c L_f i_d, \quad (3)$$
$$0 = L_f \frac{di_{d0}}{dt} + R_f i_{d0} + v_{gd0}.$$

and

$$C_f \frac{dv_{gd}}{dt} - \omega_0 C_f v_{gq} = i_d - i_{gd},$$
$$C_f \frac{dv_{gq}}{dt} + \omega_0 C_f v_{gd} = i_q - i_{gq}, \quad (4)$$
$$C_f \frac{dv_{g0}}{dt} = i_0 + i_{g0}.$$

Here, $\omega_c$ represents the fundamental angular frequency, which corresponds to the time derivative of $\theta_c$ of the grid-forming converter. We will disregard the zero sequence from this point onward. The Laplace transformation yields

$$i_d = \frac{1}{L_f s + R_f}[v_{cd} - v_{gd} + \omega_c L_f i_q],$$
$$i_q = \frac{1}{L_f s + R_f}[v_{cq} - v_{gq} - \omega_c L_f i_d]. \quad (5)$$

and

$$v_{gd} = \frac{1}{C_f s}[i_d - i_{gd} + \omega_c C_f v_{gq}],$$
$$v_{gq} = \frac{1}{C_f s}[i_q - i_{gq} - \omega_c C_f v_{gd}]. \quad (6)$$

Equations (5) and (6) are formulations representing the system plant and control requirements. This illustration highlights cross-coupling terms $(-\omega_c L_f i_d)$, $(\omega_c L_f i_q)$, $(-\omega_c C_f v_{gd})$, and $(\omega_c C_f v_{gq})$, introduced by the *abc* to *dq0* transformations. We counteract these cross-coupling effects through the incorporation of voltage and current feed-forward terms to independently design controllers for *d* and *q* [33, 34].

## III. DROOP CONTROL STRATEGY

### A. Fundamental Topology

The control unit of the grid-forming converter (Fig. 1) comprises a voltage control loop, followed by a current saturation algorithm and an inner current control loop. The outer power control loops aid in synchronization with the grid without the need for a phase-locked loop. These loops encompass active and reactive power synchronization, operating on *P–ω* and *Q–v* droop. The following sections derive mathematical representations of the generic power control loops.

### B. Active and Reactive Power Sharing

Active (*P*) and reactive power (*Q*) transmitted from the converter to the grid (Fig. 2) follow

$$P = \frac{V_c V_g}{X} \sin(\delta),$$
$$Q = \frac{V_c}{X}\big(V_c - V_g \cos(\delta)\big), \quad (7)$$

where $V_c$ is the converter voltage amplitude, $V_g$ the grid voltage, and *X* is the reactance value of the line impedance. Given the small phase angle difference $\delta$, we have, $\sin(\delta) \approx \delta$ and $\cos(\delta) \approx 1$. Consequently, Equation (7) turns into

$$\delta \approx \frac{X}{V_c V_g} P,$$
$$V_c - V_g \approx \frac{X}{V_c} Q. \quad (8)$$

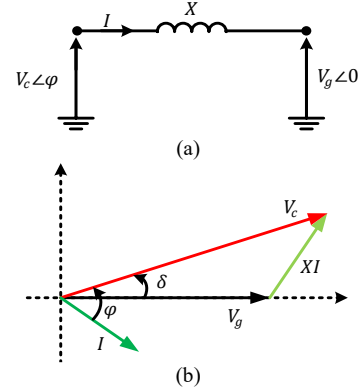

Fig. 2. (a) Equivalent circuit of a grid-forming converter. (b) Phasor diagram.

This equation indicates that the active power delivered follows the angle $\delta$, whereas the reactive power is proportional to the voltage difference $(V_c - V_g)$. Therefore, we can derive the standard droop characteristics as

$$\omega = \omega_0 + k_p(P^* - P),$$
$$v^* = v_0 + k_Q(Q^* - Q). \quad (9)$$

### C. Droop Control with Inertial Response

Conventional synchronous generators stabilize the grid through inertia, i.e., stored kinetic energy of their rotors, and suppress frequency fluctuations. As converters lack natural inertia of a rotating physical mass, we emulate the behavior of a synchronous generator and introduce virtual inertia to the grid-forming converter with low-pass behavior in the power control. Additionally, the low-pass behavior also effectively filters out unwanted measurement noise and fluctuations, reducing power oscillations and limiting fast angle variations [35] per

$$\omega_c = \omega^* + \frac{\omega_{pp}}{s + \omega_{pp}} k_p(P^* - P),$$
$$v_d^* = v^* + \frac{\omega_{pp}}{s + \omega_{pp}} k_Q(Q^* - Q). \quad (10)$$

In the specified context, $\omega^*$, $v^*$, $P^*$, and $Q^*$ define the reference values for grid angular frequency, voltage at the point of common coupling, active power reference, and reactive power reference. Additionally, $\omega_{pp}$ and $k_p$ represent the cut-off frequency of the low-pass filter and droop gains for the *P-ω* active power droop. Similarly, $\omega_Q$ and $k_Q$ signify the cut-off frequency of the low-pass filter and droop gain for the *Q–v* reactive power droop.

Fig. 3 illustrates the topology and control schematic of the grid-forming converter with the proposed current saturation algorithm. The active power droop outputs, denoted as $\omega_c$ and $\theta_c$, represent the internal angular frequency and internal angle of the converter, which we use for the Park transformation. The reactive power droop output, represented as $v_d^*$, indicates the $d$-axis voltage reference at the point of common coupling. Under normal operating conditions, we input a $d$-axis reference voltage of $v_d^* = v^*$ and a $q$-axis reference voltage of $v_q^* = 0$ at the point of common coupling.

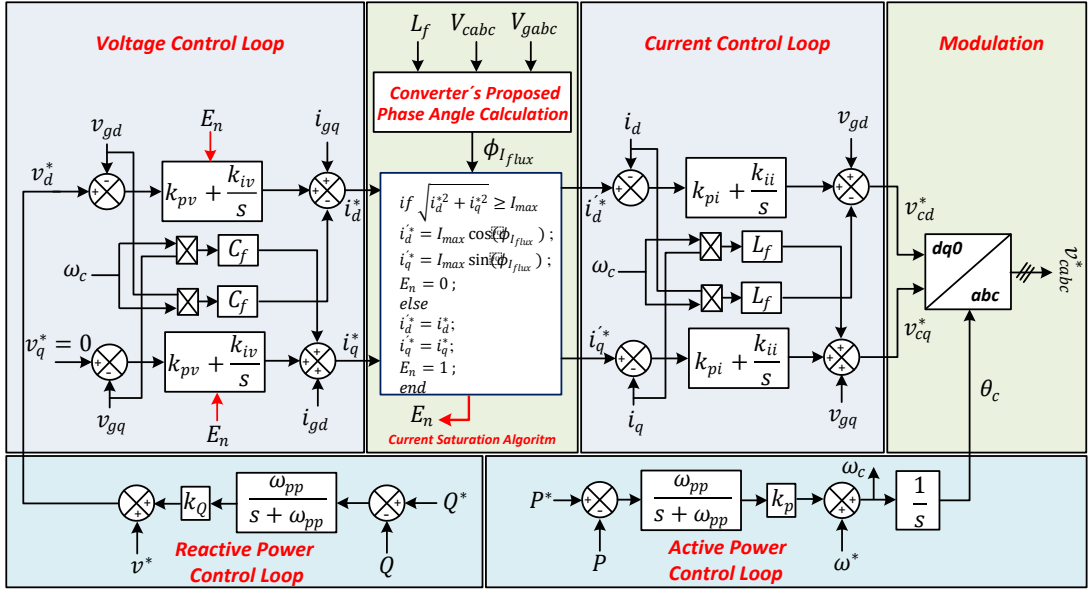

Fig. 3. Topology and control schematic diagram with the proposed current saturation algorithm of grid-forming converter.

During current saturation, the voltage controller integrator may potentially diverge. To prevent this divergence, we implement anti-windup feedback during the saturation mode. In this scenario, the saturation algorithm sends a feedback signal ($E_n = 0$) to the controller integrator of the voltage PI controller, as depicted in Fig 4. Conversely, during normal operation, the feedback signal remains active ($E_n = 1$).

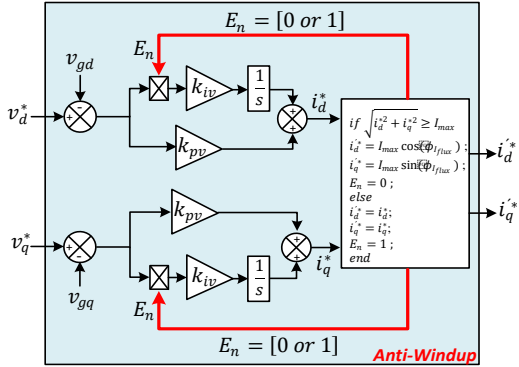

Fig. 4. Structure of the current saturation algorithm and anti-windup.

## IV. CURRENT SATURATION METHODS
### A. Current Reference Saturation

A straightforward approach to limit current involves separately saturating the $d$- and $q$-components of the current, as presented in the literature [36]. The designer must choose the maximum current components $I_{dq}^*$ so that the maximum current amplitude $I_{max}$ always remains below the specified threshold. The relationship among these three parameters is given by

$$I_{max} = \sqrt{I_{dmax}^2 + I_{qmax}^2}. \tag{11}$$

If only one component reaches its maximum value, the current does not necessarily assume its maximum value. Additionally, in such scenarios, the current references shift to a different angle before saturation results in a mismatch of the desired power factor imposed by the current references on the converter. Both issues are discussed graphically in the literature [36], where the current reference reaches its maximum value $I_{dmax}$ and becomes saturated.

### B. Saturation of Current Amplitude

This method solves the two drawbacks outlined in the preceding paragraph. It requires both the current reference amplitude $|I_{dq}^*|$ and the current phase angle $\phi$ derived from the $dq$-components per

$$|I_{dq}^*| = \sqrt{I_d^{*2} + I_q^{*2}}, \tag{12}$$

and

$$\phi = \tan^{-1}\left(\frac{I_q^*}{I_d^*}\right). \tag{13}$$

The method directly constrains the amplitude and requires the angle to compute the saturated values of each current component, as

$$\begin{cases} i_{dq}^* = i_{dq}^* & \text{if } \sqrt{i_d^{*2} + i_q^{*2}} < i_{max}^{sat}, \\ i_d^* = i_{max}^{sat}\cos(\phi) & \text{if } \sqrt{i_d^{*2} + i_q^{*2}} \geq i_{max}^{sat}, \\ i_q^* = i_{max}^{sat}\sin(\phi) & \text{if } \sqrt{i_d^{*2} + i_q^{*2}} \geq i_{max}^{sat}. \end{cases} \tag{14}$$

The currents $i_d^*$, $i_q^*$, represent the input reference for the current control loop, and $i_{max}^{sat}$ is the maximum saturated current. These values are derived from the output of the voltage control loop

## V. TRANSIENT STABILITY OF GRID-FORMING CONVERTER BASED ON DROOP CONTROL

### A. Transient Stability Analysis of a Droop-controlled Grid-Forming Converter without a Current Saturation Algorithm

The simplified quasi-static model (Fig. 2) is based on the assumptions presented in Section 3. This initial approach does not incorporate the current-saturation algorithm. The simplified model consolidates real electrical parameters such as $I$, $V_c$, and the AC voltage $V_g$. Therefore, during quasi-static operation, the active power can be computed using the formula $P = P_{max} \sin(\delta)$, where $P_{max} = \frac{V_c V_g}{X}$ and $\delta$ represents the angular difference between the grid-forming converter voltage $V_c$ and the grid AC voltage $V_g$.

Starting from the simplified model (15) as a basis, Fig. 5 illustrates the synchronization procedure of the grid-forming converter during a three-phase bolted fault at the denoted point (F). Within this context, for a specified value of $P^*$, two equilibrium points (denoted as (01) and (01')) are present, where the power $P$ matches its reference $P^*$. However, it's noteworthy that only equilibrium point (01) is considered stable in small-signal analysis [37]. According to (15), the initial angle $\delta_0$ is determined by

$$\delta_0 = \sin^{-1}\left(\frac{P^*}{P_{max}}\right). \tag{15}$$

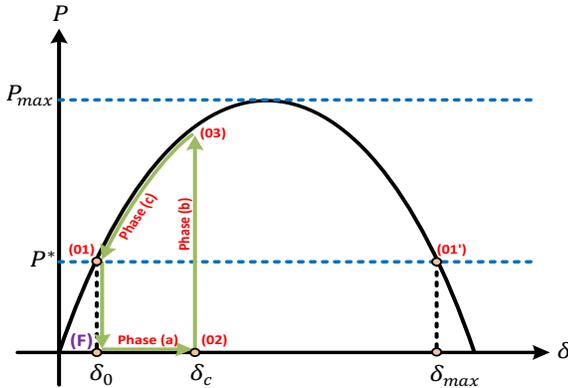

Fig. 5. $P(\delta)$ diagram of a droop control-based grid-forming converter without current saturation.

The maximum angle $\delta_{max}$ of the operational point (01') is formulated as

$$\delta_{max} = \pi - \delta_0. \tag{16}$$

Based on (9), in normal operation, $\omega_c = \omega_0$. When the fault happens, $P$ immediately drops to zero. This drop causes an instantaneous deviation in frequency, which is expressed as

$$\omega_c = k_p P^* + \omega_0. \tag{17}$$

In Fig. 5, as the fault occurs, the converter angular frequency $\omega_c$ exceeds the grid angular frequency $\omega_g$ (hyper-synchronism), which causes an increase in the angle (Phase (a)). Following fault clearance, the voltage is supposed to promptly recover. Consequently, a vertical line extends directly from operating point (02) to (03) (Phase (b)). Throughout this phase, $\delta$ equals $\delta_c$ (where $\delta_c$ is termed the clearing angle). Towards the end of Phase (b), $P$ surpasses $P^*$, resulting in $\omega_c < \omega_g$ (hypo-synchronism). It leads to a decrease in the angle from (03) back to the equilibrium point (01) (Phase (c)), wherein $\omega_c$ equals $\omega_g$ and $P$ equals $P^*$. The stability threshold is reached upon the clearing angle attaining $\delta_{max}$, which denoted as the critical clearing angle $\delta_{cc}$ [19]. Therefore, the system is under a risk of stability loss if the fault time is long.

### B. Transient Stability Analysis of a Droop-Controlled Grid-Forming Converter with Current Saturation

We incorporate a current saturation algorithm to integrate current saturation into this analysis. When the current exceeds its maximum allowable magnitude ($I_{max,sat}$) while synchronized on the $d$-axis, the expression for the delivered active power becomes

$$P = P_{max,sat} \cos(\delta), \tag{18}$$

where $P_{max,sat} = V_g I_{max,sat}$.

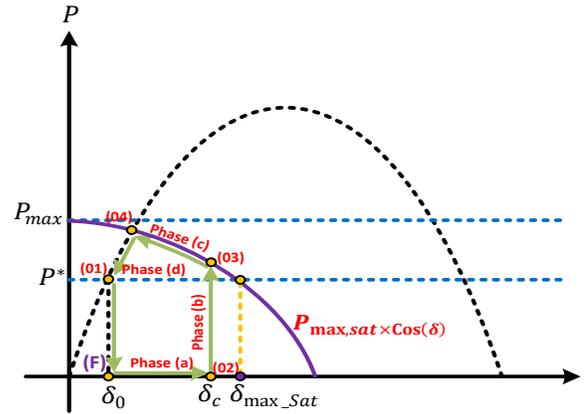

Fig. 6. $P(\delta)$ diagram of a droop control-based grid-forming converter with current saturation.

Fig. 6 illustrates the progression of the angle in the current-saturated mode. The sequence of operation is now as follows:
- In Phase (a), whether or not the current saturation algorithm is implemented, the behavior remains the same.
- Phase (b) involves fault clearance, resulting in the transition of the operating point to (03), now positioned within the curve represented by $P = P_{max,sat} \cos(\delta)$.
- During Phase (c), the operating point shifts from (03) to (04). This intersection indicates that the normal operating voltage can be restored, thus facilitating a smooth transition from saturated to unsaturated mode.
- Phase (d) sees the system transitioning from (04) back to its equilibrium point (01).

The implementation of the current saturation algorithm reduces $P_{max}$ to $P_{max,sat}$. Consequently, the system maintains stability if the fault is rectified prior to reaching $\delta_{max,sat}$ as defined in

$$\delta_{max,sat} = \cos^{-1}\left(\frac{P^*}{P_{max,sat}}\right). \tag{19}$$

Based on this equation, we now set the stability limit of the system as $\delta_{cc} = \delta_{max,sat}$.

The absence of a current saturation algorithm increases the risk of instability during extended faults, potentially leading to voltage instability and cascading failures. However, implementing a current saturation algorithm helps maintain system stability by controlling current levels and ensuring a prompt return to normal voltage conditions after faults, which also improves overall operational reliability and efficiency in grid-forming converters.

## VI. PROPOSED CURRENT-SATURATION ALGORITHM ANALYSIS

Previous studies conventionally set the angle $\phi$, which represents the current phase angle, to zero. They then calculated this angle directly based on the $I_d$ and $I_q$ current references during the operation of current saturation [20, 38]. However, introducing a novel calculation of this angle as a control parameter in the current saturation algorithm achieves significant improvements in system fault ride-through and grid fault support with transient synchronization stability during faults.

### A. Virtual Flux Estimation for Grid Synchronization

The concept of virtual flux builds on the definition of *flux* ($\psi$), which integrates voltage ($V$) as per (20). This approach is inspired by converter and electric drive system control [32] per

$$\psi = \int V dt + \psi_0. \tag{20}$$

We estimated the virtual flux $\psi_g$ at the grid side of the filter inductor ($R_f, L_f$) based on the converter's output voltage. We derive it from the PWM reference signals and the measured DC-link voltage. This estimation replaces the direct voltage measurements at the grid side of the filter inductor in our system.

The equivalent circuit corresponding to Fig. 2 allows the representation of system voltages by

$$\vec{V}_c = jX\vec{I} + \vec{V}_g. \tag{21}$$

### B. Dynamic Equations

The electrical equations associated with each of the three phases of the grid-forming converter's output filter follow

$$v_{ck} = L_f \frac{di_k}{dt} + R_f i_k + v_{gk}, \tag{22}$$

where the index *k* corresponds to the phases *a, b, c*.

We define the voltages $v_{gk}$ as the derivative of a virtual flux $\psi_{gk}$ at the terminals of the grid-forming converter as follows

$$v_{gk} = \frac{d\psi_{gk}}{dt}. \tag{23}$$

By substituting (23) into (22), organizing the derivative terms, and treating $L_f$ as a constant, (22) turns into

$$v_{ck} = L_f \frac{di_k}{dt} + R_f i_k + \frac{d\psi_{gk}}{dt}, \tag{24}$$

which leads to

$$v_{ck} = R_f i_k + \frac{d}{dt}(L_f i_k + \psi_{gk}). \tag{25}$$

The latter entails the derivative of the virtual flux for each phase of the converter as

$$\psi_{ck} = L_f i_k + \psi_{gk}. \tag{26}$$

For each phase, we calculate the converter virtual flux, $\psi_{ck}$, by summing the instantaneous current, $i_k$, multiplied by $L_f$, and the grid side virtual flux, $\psi_{gk}$, measured at the output terminals of the grid-forming converter. Direct measurement of voltages and currents (Fig. 1) allows for the direct calculation of $\psi_{ck}$ according to (26). We find that parameter variations largely do not affect the accuracy of estimating $\psi_{ck}$ due to the stable nature of the filter inductance $L_f$.

Fig. 7 illustrates the estimation outlined in (21) and the relationship between voltage and virtual flux vectors and assumes negligible resistance in the filter inductor. According to (21) and Fig. 7 the internal voltage $V_c$ is proportional to the virtual flux $\psi_c$ and therefore, its control allows to regulate the active and reactive power exchanged by the grid-forming converter. Furthermore, the angle of the voltage vector ($\vec{V}_g$) serves as the angular reference. The current vector lags by an angle ($\varphi$), indicating power delivery from the grid-forming converter to the grid. The internal voltage ($\vec{V}_c$) is calculated according to (21) with its angular position corresponding to a positive power angle ($\delta > 0$) in addition to a higher magnitude than $\vec{V}_g$ ($|\vec{V}_c| > |\vec{V}_g|$). We express the virtual flux vector $\vec{\psi}_c$ as $\vec{\psi}_c = L_f \vec{I} + \vec{\psi}_g$, where $\vec{\psi}_c$ leads $\vec{\psi}_g$ by an angle $\delta$ and has a greater magnitude. This relationship demonstrates that the converter can generate both active and reactive power freely, similar to how a synchronous generator supplies active and reactive power to the grid [30, 39].

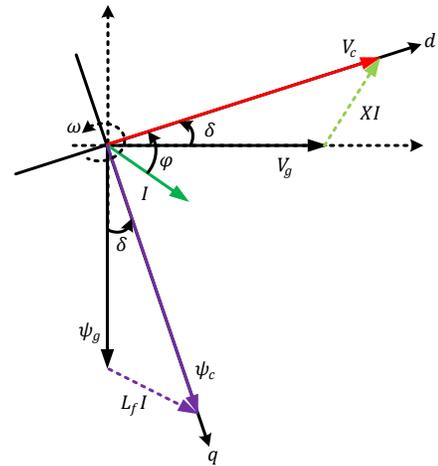

Fig. 7. Vector diagram of the grid-forming converter in steady-state operation.

## VII. PROPOSED CURRENT PHASE ANGLE FOR CURRENT-SATURATION ALGORITHM

The proposed current-saturation algorithm follows the previous analysis outlined in Section VI to estimate the virtual fluxes at the output of the converter and the grid sides. Per (26), we use these estimated fluxes to compute the current of the filter inductor. Moreover, the calculation of the virtual flux entails integrating the converter and grid voltages and transforming them into $d$ and $q$ components using Park transformation. The integration process inherent in virtual flux estimation induces a filtering effect, generating a filtered reference currents resulting in a stable and filtered current phase angle that contributes to system synchronization stability and enhances transient performance during grid faults.

The converter virtual flux follows from (26) after the incorporation of the $d$ and $q$ components of the converter's currents and voltages per

$$\psi_{cd} = L_f i_d + \psi_{gd},$$
$$\psi_{cq} = L_f i_q + \psi_{gq}. \quad (27)$$

The reference currents follow as

$$I_{df} = \frac{1}{L_f}(\psi_{cd} - \psi_{gd}),$$
$$I_{qf} = \frac{1}{L_f}(\psi_{cq} - \psi_{gq}). \quad (28)$$

We then estimate the current phase angle per

$$\phi_{I_{flux}} = \tan^{-1}\left(\frac{I_{qf}}{I_{df}}\right). \quad (29)$$

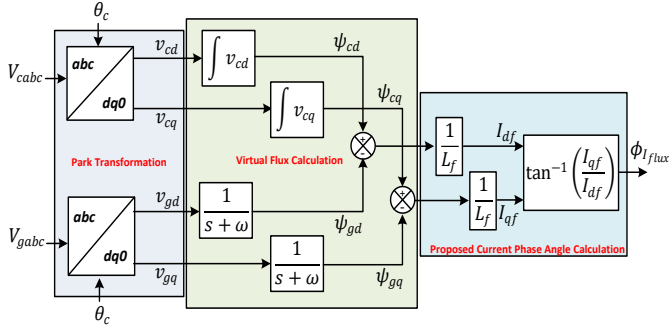

Fig. 8. Block diagram of the proposed current phase angle.

To avoid derivatives and unknown initial conditions, we use a first-order filter to approximate the integral for the virtual flux of the grid voltages. A filter cut-off frequency close to 1 Hz minimizes errors in the low-frequency range during virtual-flux estimation, especially at the fundamental frequency (50/60 Hz) [30].

### A. Proposed Current-Saturation Algorithm

The conventional current saturation of (14) and the proposed current phase angle from (29) provides a new current-saturation algorithm with saturated current references as

$$\begin{cases} i_{dq}^{'*} = i_{dq}^{*} & \text{if } \sqrt{i_d^{*2} + i_q^{*2}} < i_{max}^{sat}, \\ i_d^{'*} = i_{max}^{sat}\cos(\phi_{I_{flux}}) & \text{if } \sqrt{i_d^{*2} + i_q^{*2}} \geq i_{max}^{sat}, \\ i_q^{'*} = i_{max}^{sat}\sin(\phi_{I_{flux}}) & \text{if } \sqrt{i_d^{*2} + i_q^{*2}} \geq i_{max}^{sat}, \end{cases} \quad (30)$$

where the current phase angle $\phi_{I_{flux}} = \tan^{-1}\left(\frac{I_{qf}}{I_{df}}\right)$.

TABLE I
SYSTEM AND CONTROL PARAMETERS

| System Parameters | | |
|---|---|---|
| Description | Symbol | Parameter |
| Grid-forming converter filter inductance | $L_f$ | 2 mH |
| Grid-forming converter filter resistance | $R_f$ | 1 mΩ |
| Grid-forming converter filter capacitance | $C_f$ | 60 μF |
| Grid filter inductance | $L_g$ | 8 mH |
| Grid filter resistance | $R_g$ | 100 mΩ |
| Point of common coupling loads | $R_{Labc}$ | 10Ω |
| Grid-forming converter power rating | $P_n$ | 30 kW |
| Nominal AC voltage | $V_n$ | 480 V |
| Nominal DC voltage | $V_{dc}$ | 1 kV |
| Nominal AC angular frequency | $\omega_n$ | 314 rad/s |
| Switching frequency | $f_s$ | 10 kHz |
| Maximum saturated current | $i_{max}^{sat}$ | 110 A |
| Control Parameters | | |
| Cut-off frequency of low pass filter | $\omega_{pp}$ | 35 rad/s |
| Active power droop coefficient | $k_P$ | 0.001 |
| Reactive power droop coefficient | $k_Q$ | 0.001 |
| Voltage integrator regulators gains | $k_{iv}$ | 500 |
| Voltage proportional regulators gains | $k_{pv}$ | 0.55 |
| Current integrator regulators gains | $k_{ii}$ | 5000 |
| Current proportional regulators gains | $k_{Pi}$ | 500 |

## VIII. RESULTS AND DISCUSSION

### A. Simulation Results

Figs. 9, 10, and 11 present the simulation results of the estimated current phase angle based on the conventional and the proposed method under various faults: (*i*) three-phase 100% voltage sag, (*ii*) two-phase short circuits to ground, and (*iii*) three-phase faults resulting in a 50% phase shift. The proposed current phase angle has a significantly smoother response than the conventional approach. Consequently, the filtered current phase angle enables stable and reliable $I_d^*$ and $I_q^*$ reference currents for the PI current controller. The current saturation algorithm presented here facilitates smooth and stable transient synchronization of the system post-fault, surpassing the performance of the conventional method.

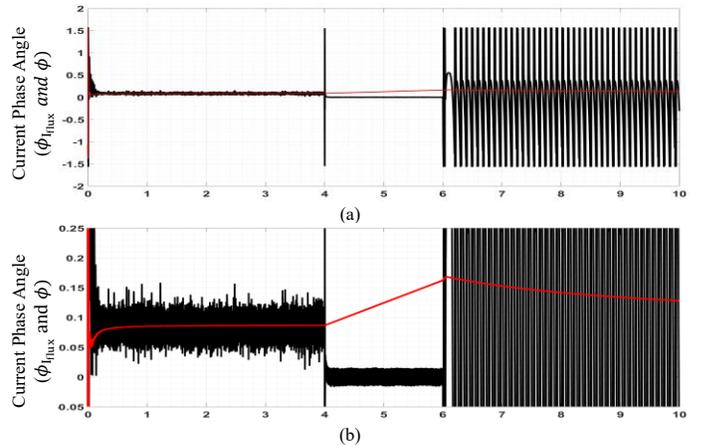

Fig. 9. Three-phase 100% voltage sag results of conventional ($\phi$, black line) and proposed ($\phi_{I_{flux}}$, red line) angles: (a) current phase angle, (b) zoomed figure of current phase angle.

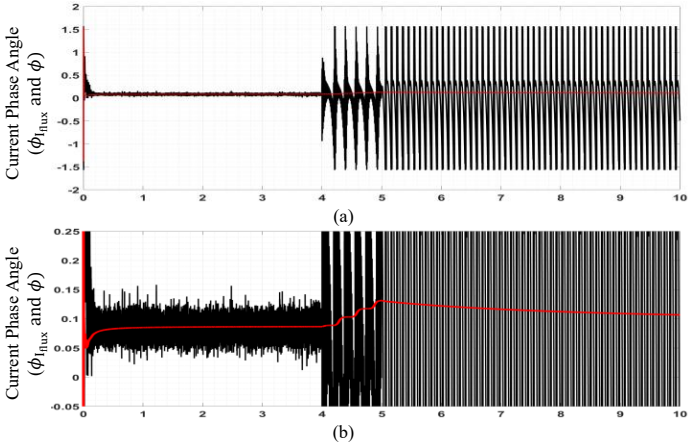

Fig. 10. Two phases short circuit with the ground of conventional ($\phi$, black line) and proposed ($\phi_{I_{\text{flux}}}$, red line) angle: (a) current phase angle, (b) zoomed figure of current phase angle.

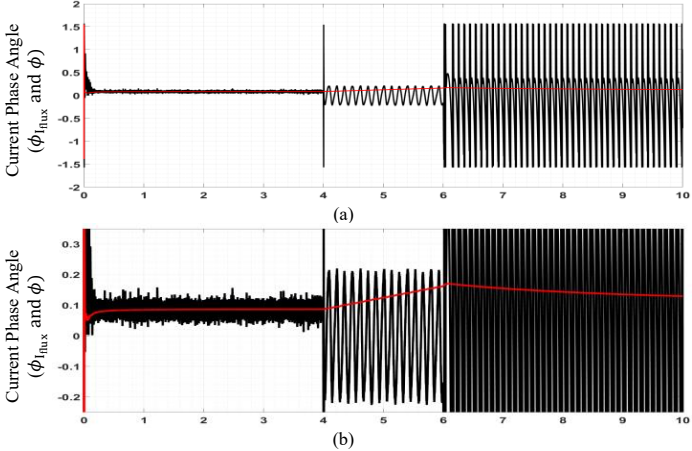

Fig. 11. Three-phase 50% phase shift of conventional ($\phi$, black line) and proposed ($\phi_{I_{\text{flux}}}$, red line) angle: (a) current phase angle, (b) zoomed figure of current phase angle.

### B. Hardware-In-The-Loop Results

We analyzed and compared the performance of the novel current saturation method with the state of the art in a hardware-in-the-loop setup (RT-LAB OP4610, Fig. 12) [40]. Table I lists the system and control parameters. Additionally, we examined and validated the system's performance and dynamics particularly under the following grid faults:

- A 5 s duration of 100% three-phase grid voltage sag;

- A 3.5 s duration of two-phase short circuit with ground;

- A 5 s duration of 50% three-phase shift.

(*i*) *100% Three-Phase Grid Voltage Sag*

We introduce a 5 s voltage sag at the point of common coupling (Fig. 13(a) for the conventional current saturation algorithm, Fig. 14(a) for the new one). When this fault occurs, the control system swiftly activates the current saturation in response to the AC voltage drop to protect the converter ($i_{\text{max}}^{\text{sat}} = 110$ A).

Figs. 14(a) and 14(b) demonstrate that the proposed current saturation algorithm allows the system to recover smoothly and to swiftly return to its equilibrium point. In Figs. 13(a) and 13(b), however, the conventional method loses synchronicity and fails to recover stability when confronted with the same grid faults.

Subfigures (c)–(e) of Figs. 13 and 14 provide a detailed comparison of the output grid-forming converter current behavior under both the conventional and proposed current saturation algorithms. In Fig. 13(c), the conventional method exhibits instability following the grid fault, whereas the proposed method in Fig. 14(c) does not.

Figs. 13(f), 13(h), and their zoomed versions 13(g), 13(i), display the PI controller's inner loop direct current, its reference, and the grid-forming converter's output active power with grid frequency for the conventional current saturation algorithm. These traces highlight the system's failure to restore its stability and normal operation after the fault is cleared. In contrast, their equivalents in Figs. 14(f)–(i) illustrate the smooth stabilization and recovery achieved with the proposed current saturation algorithm.

(*ii*) *Two-Phase Short Circuit with Ground Fault*

Figs. 15 and 16 demonstrate the two systems' response to a two-phase short circuit to ground lasting 3.5 s. After we cleared the grid fault, the proposed current saturation algorithm recovers smoothly and rapidly (Figs. 15(a)–(d)). Again, the conventional current saturation method loses synchronization and struggled to regain stability post-fault (Figs. 16(a)–(d)).

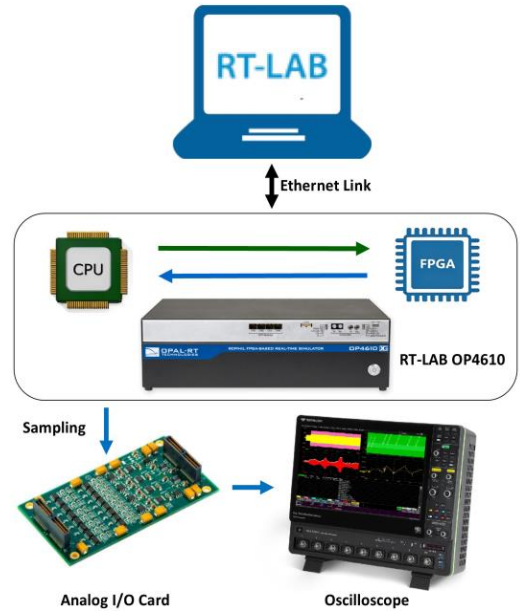

Fig. 12. Hardware-in-the-loop setup.

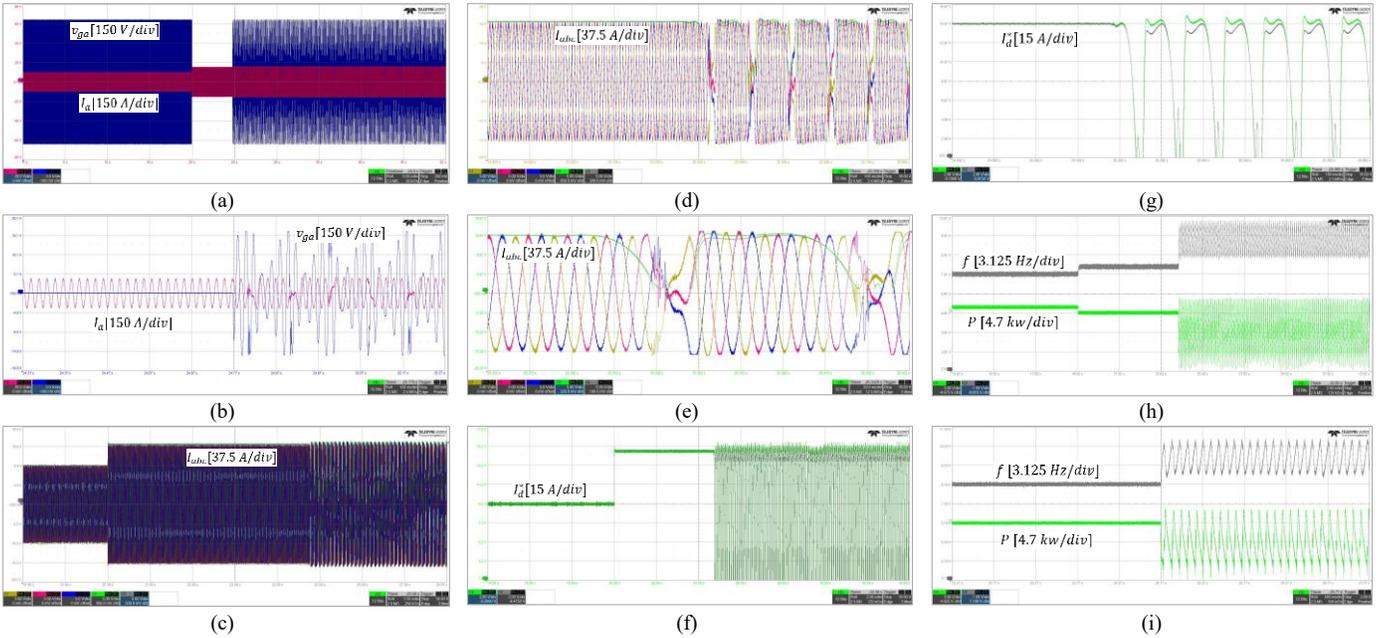

Fig. 13. The conventional current saturation algorithm during a 100% three-phase voltage sag for the grid-forming converter: (a) PCC voltage and output current, (b) zoomed-in view of PCC voltage and output current, (c) three-phase currents, (d) zoomed-in view of three-phase currents, (e) further zoomed-in view of three-phase currents, (f) $d$-axis controller current and its reference, (g) zoomed-in view of $d$-axis controller current and its reference, (h) grid frequency and output active power, (i) zoomed-in view of grid frequency and output active power.

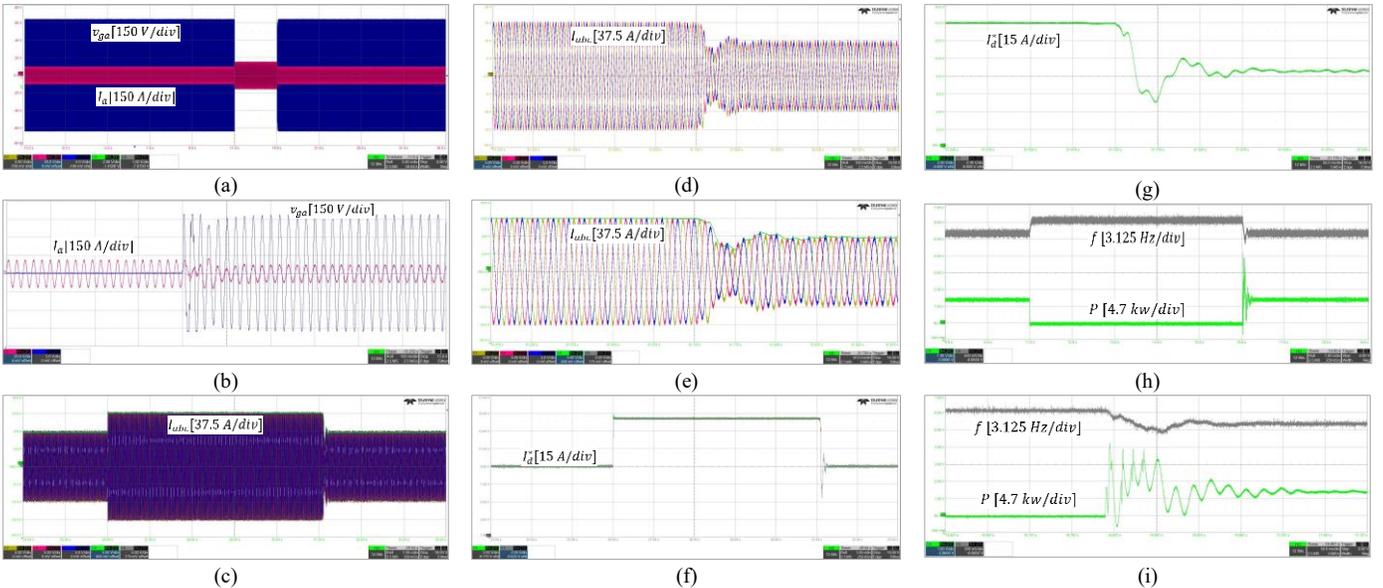

Fig. 14. The proposed current saturation algorithm during a 100% three-phase voltage sag for the grid-forming converter: (a) PCC voltage and output current, (b) zoomed-in view of PCC voltage and output current, (c) three-phase currents, (d) zoomed-in view of three-phase currents, (e) further zoomed-in view of three-phase currents, (f) $d$-axis controller current and its reference, (g) zoomed-in view of $d$-axis controller current and its reference, (h) grid frequency and output active power, (i) zoomed-in view of grid frequency and output active power.

A closer look at the inner loops of the PI controllers of both methods (Subfigures (e) and (f)) illustrate the challenges faced by the conventional system in restoring stability and normal operation post-fault.

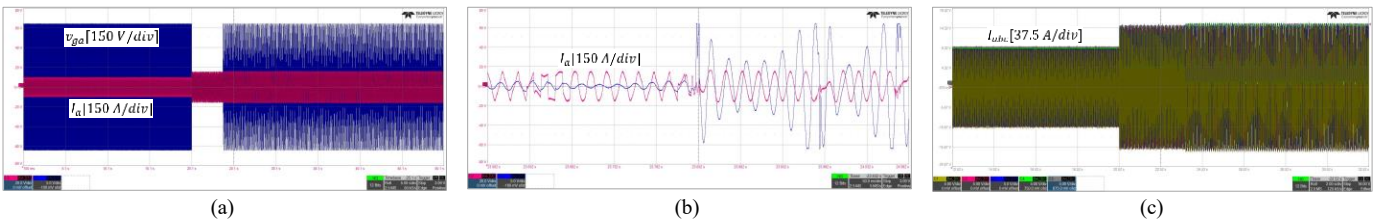

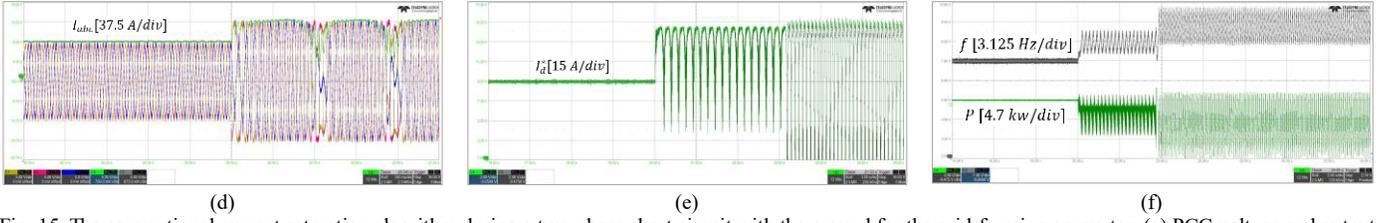

Fig. 15. The conventional current saturation algorithm during a two-phase short circuit with the ground for the grid-forming converter: (a) PCC voltage and output current, (b) zoomed-in view of PCC voltage and output current, (c) three-phase currents, (d) zoomed-in view of three-phase currents, (e) *d*-axis controller current and its reference, (f) grid frequency and output active power.

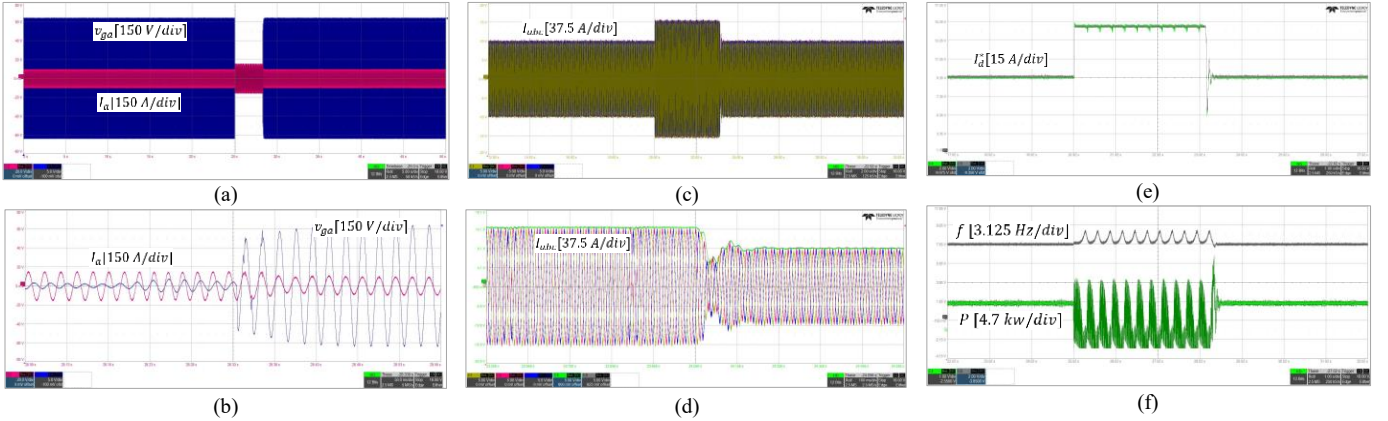

Fig. 16. The proposed current saturation algorithm during a two-phase short circuit with the ground for the grid-forming converter: (a) PCC voltage and output current, (b) zoomed-in view of PCC voltage and output current, (c) three-phase currents, (d) zoomed-in view of three-phase currents, (e) *d*-axis controller current and its reference, (f) grid frequency and output active power.

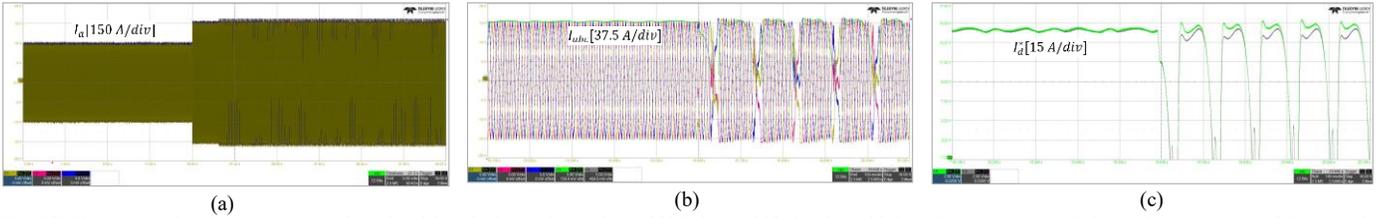

Fig. 17. The conventional current saturation algorithm during a three-phase 50% phase shift for the grid-forming converter: (a) three-phase currents, (b) zoomed-in view of three-phase currents, (c) *d*-axis controller current and its reference.

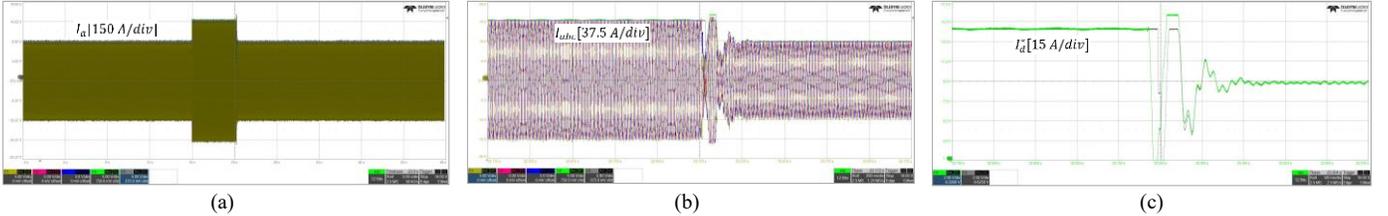

Fig. 18. The proposed current saturation algorithm during a three-phase 50% phase shift for the grid-forming converter: (a) three-phase currents, (b) zoomed-in view of three-phase currents, (c) *d*-axis controller current and its reference.

(*iii*) *Three-Phase 50% Phase-Shift Fault*

We conducted the last test to evaluate the efficiency and performance of the system under a 50% three-phase shift fault. Fig. 17 and Fig. 18 illustrate the respective responses of the system using the conventional and the proposed current saturation algorithm. As for the other test cases, as seen in Fig. 18, the proposed algorithm demonstrates a swift recovery after the fault clearance without any spikes or transient divergence. The conventional algorithm in contrast falls into a limit-cycle oscillation post-fault, which the zoomed version of the d/q current in Fig. 17 shows with more detail.

## IX. CONCLUSION

The change from voltage- to current-source mode in grid-forming converters poses a risk to synchronization stability in combination with current saturation. In response to that challenge, this paper presents a novel current saturation algorithm aimed at enhancing fault ride-through performance and synchronization stability. We introduce a new calculation method for the saturated current references based on virtual fluxes as new control parameters for the current phase angle and demonstrate the crucial significance for maintaining the synchronization. We particularly calculate the saturated-current phase angle with a virtual-flux method. Furthermore, our method optimizes the distribution of active and reactive currents to control stability of grid-forming converters during grid faults. We compared the method with conventional current saturation in various fault scenarios, where the conventional method is known to struggle. In contrast to the state of the art, the new saturation method demonstrates rapid system recovery and keeps the system synchronized even under challenging conditions. Accordingly, our contribution could solve a concrete problem that hampers the translation of grid-forming into real-world use to stabilize future weak and/or low-inertia grids.